\documentclass[epj]{svjour}

\usepackage{graphicx}
\usepackage{epsfig}
\usepackage{amsfonts}
\usepackage{amsmath,amssymb}

\begin{document}

\title{Shell-structure fingerprints of tensor interaction}

\author{M. Zalewski\inst{1} \and W. Satu{\l}a\inst{1}\inst{2}
\and J. Dobaczewski\inst{1}\inst{3} \and P. Olbratowski\inst{1}
\and M. Rafalski\inst{1} \and T. R. Werner\inst{1} \and R. A. Wyss\inst{2}
%
%
}                     
%
%
\institute{Institute of Theoretical Physics, University of Warsaw, ul. Ho\.za
69, 00-681 Warsaw, Poland \and
KTH (Royal Institute of Technology),
AlbaNova University Center, 106 91 Stockholm, Sweden
\and
Department of Physics, P.O. Box 35 (YFL),
FI-40014 University of Jyv\"askyl\"a, Finland
}

\date{Received: \today / Revised version: \today}

\abstract{We address consequences of strong tensor and weak
spin-orbit terms in the local energy density functional, resulting
from fits to the $f_{5/2} - f_{7/2}$ splittings in $^{40}$Ca,
$^{48}$Ca, and $^{56}$Ni. In this study, we focus on nuclear binding
energies. In particular, we show that the tensor contribution to the
binding energies exhibits interesting topological features closely
resembling that of the shell-correction. We demonstrate that in the
extreme single-particle scenario at spherical shape, the tensor
contribution shows tensorial magic numbers equal to $N(Z)$=14, 32,
56, and 90, and that this structure is smeared out due to
configuration mixing caused by pairing correlations and migration of
proton/neutron sub-shells with neutron/proton shell filling.  Based
on a specific Skyrme-type functional SLy4$_T$, we show that the
proton tensorial magic numbers shift with increasing neutron excess
to $Z$=14, 28, and 50.}

\PACS{
{21.60.Jz}{Nuclear Density Functional Theory and extensions (includes Hartree-Fock
and random-phase approximations)}
{21.60.-n} {Nuclear structure models and methods}
}
\maketitle

\section{Introduction}
\label{sec1}

Density functional theory (DFT) is a universal
{\it ab initio\/} approach designed and used to
calculate properties of electronic systems entrapped in the external Coulomb field
of nuclei. It has been successfully applied to
atoms, molecules, or condensed media. Universality of the DFT
means independence of a functional form of the shape of external one-body
potential holding the electronic system together. The existence of such a
universal and, in principle, exact density functional describing
ground-states of externally bound fermionic systems is guaranteed by the
Hohenberg-Kohn~\cite{[Hoh64]} and Kohn-Sham~\cite{[Koh65a]} (HKS) theorems.

Generalization of the DFT theory to self-bound systems like atomic nuclei
encounters problems associated with intrinsic rather than
laboratory density which characterizes the atomic nuclei, see
Ref.~\cite{[Eng07],[Gir08],[Gir08a]}. In spite of that, the HKS
theorem have strongly influenced the way of thinking in the field of nuclear
structure. Nowadays, the nuclear structure theorists
employ the functionals that are treated as separate entities,
independently, to a large extent, of the underlying effective nucleon-nucleon (NN)
interactions like, for example, the local Skyrme interaction~\cite{[Sky56xw]}.
Free parameters of these functionals are directly adjusted to fit empirical
data. There are also attempts to enrich their functional form as compared
to the form resulting from the mean-field (MF) averaging of the effective
Skyrme interaction~\cite{[Car08]}, which are motivated by a rather mediocre
performance of the conventional Skyrme-type functionals, see for example Refs.~\cite{[Kor08],[Klu08]}.

Adequateness of the fitting strategy -- that is, the choice of the data set -- is a key
factor determining performance of the nuclear energy density functional (EDF)
method. In this work we explore the Skyrme-type local EDF
approach to nuclear structure and focus on the spin-orbit
(SO) and tensor parts of the functional. Throughout the years, not much
attention was paid to, in particular, the tensor part, mainly due to the lack of
clear experimental data constraining the strength of this part of the EDF.
Hence, the tensor term in the existing Skyrme functionals is
either trivially set to zero by hand, see the review in Ref.~\cite{[Ben03]}, or is a result of a global fit
to bulk nuclear properties \cite{[Les07]}.

\begin{figure}[!]
\includegraphics[width=0.4\textwidth, clip]{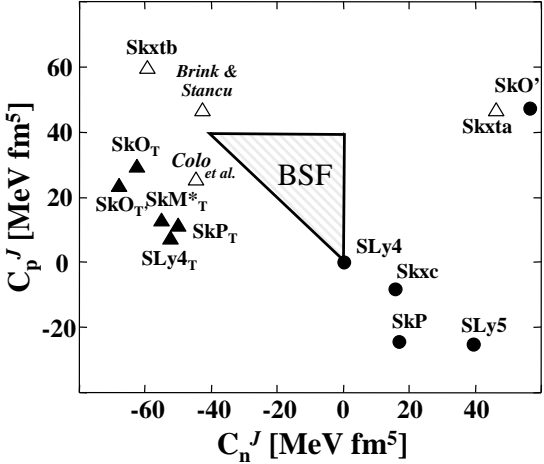}
\caption{Proton $C_p^J = (C_0^J -C_1^J)/2$
versus neutron $C_n^J = (C_0^J +C_1^J)/2$ tensor coupling
constants resulting form fits to: bulk nuclear properties
(black dots) and the s.p. levels and the SO splittings (triangles).
Open triangles represent fits of Ref.~\cite{[Bro06],[Col07],[Bri07w]}.
Black triangles mark our results~\cite{[Zal08],[Sat08],[Zal08as]} from
the fit to the $f_{7/2}-f_{5/2}$ SO splittings. Shaded area represents
the parameters established by Brink, Stancu and Flocard (BSF) in their seminal
paper~\cite{[Bri77]}
}.
\label{fig1}
\end{figure}

Recent revival of interest in the tensor term was
triggered by systematic observation of non-conventional shell evolution
in isotopic chains of light nuclei far from stability, including
new magic shell-gap opening at $N$=32,
see for example Refs.~\cite{[For04],[Din05]}. Otsuka and
collaborators associated these effects with the two-body tensor
interaction~\cite{[Ots05],[Ots06]}. This interpretation was soon confirmed
within the local Skyrme-type EDF models~\cite{[dob06c],[Bro06],[Dob07b],[Sug07a],[Col07],[Bri07w]}.
Inclusion of single-particle (s.p.) energies
in the fitting data sets appears to lead to the tensor coupling
constants~\cite{[Bro06],[Col07],[Bri07w],[Zal08],[Sat08],[Zal08as]} which
are at variance with bulk fits, see Fig.~\ref{fig1} and extensive
discussion in Ref.~\cite{[Les07]}.

The aim of this work is to look into
consequences of strong attractive isoscalar and isovector tensor fields and
weak SO fields resulting from the fitting method proposed by our
group~\cite{[Zal08]}. In particular, we study such consequences for the nuclear binding energies.
The paper is organized as follows. In Sec. \ref{sec2},
theoretical framework is briefly outlined. In Sec. \ref{sec3}, the procedure
used to fit the tensor and SO coupling constants is discussed. In Sec.~\ref{sec4},
numerical results showing tensor energy contribution to the total
binding energy, followed by a discussion of {\it tensorial magic structure\/},
is presented. The paper is summarized in Sec.~\ref{summary}.

\vspace{0.5cm}

\section{Tensor and spin-orbit parts of the local nuclear
energy density functional}
\label{sec2}

In this work we consider the local EDF
${\mathcal H}({\mathbf r})$ of  the Skyrme-type.
It consists of a kinetic energy and a sum of isoscalar ($t$=0) and isovector
($t$=1) potential energy terms:
\begin{equation}\label{eq108}
   {\mathcal H}({\mathbf r}) = \frac{\hbar^2}{2m}\tau_0
               + \sum_{t=0,1} \biggl\{ {\mathcal H}_t({\mathbf
               r})^{\text{even}} + {\mathcal H}_t ({\mathbf r})^{\text{odd}}
               \biggr\} ,
   \end{equation}
where
\begin{eqnarray}
\label{hte}
\mathcal{H}_t^{\text{even}}
& = & C^{\rho}_t[\rho_0] \rho^2_t + C^{\Delta \rho}_t
\rho_t\Delta\rho_t + \\ \nonumber
&\quad & C^{\tau}_t
\rho_t\tau_t + C^J_t {\mathbb J}^2_t +
C^{\nabla J}_t \rho_t {\mathbf \nabla}\cdot{\mathbf  J}_t,
\end{eqnarray}
\begin{eqnarray}
\label{hto}
\mathcal{H}_t^{\text{odd}} & = &C^{s}_t [\rho_0 ] {\mathbf s}^2_t
+ C^{\Delta s}_t {\mathbf s}_t\cdot\Delta {\mathbf s}_t +  \\ \nonumber
&\quad &C^{T}_t{\mathbf s}_t \cdot {\mathbf T}_t +
C^j_t {\mathbf j^2_t} +
C^{\nabla j}_t {\mathbf s}_t \cdot ({\mathbf \nabla}\times {\mathbf j}_t),
\end{eqnarray}
with the density-dependent primary coupling constants $C^{\rho}_t[\rho_0]$
and $C^{s}_t [\rho_0 ]$. The potential energy terms are bilinear forms of
either time-even ($\rho$, $\tau$, ${\mathbb J}$)
or time-odd (${\mathbf s}$, ${\mathbf j}$, ${\mathbf T}$) densities
and their derivatives, see, e.g.. Ref.~\cite{[Ben03]} for details.
The ${\mathbf J}_{t}$ density denotes the vector part of the spin-current
tensor, ${\mathbf J}_{t,\lambda} = \sum_{\mu\nu}
\epsilon_{\lambda\mu\nu}{\mathbb J}_{t,\mu\nu}$.

In this work we focus on the tensor,
\begin{equation}
\label{htta}
\mathcal{H}^T     =  C^J_0 {\mathbb J}^2_0 +
                       C^J_1 {\mathbb J}^2_1 ,
\end{equation}
and the SO terms,
\begin{equation}
\label{httb}
\mathcal{H}^{SO}  =  C^{\nabla J}_0 \rho_0 {\mathbf \nabla}\cdot{\mathbf J}_0  +
                       C^{\nabla J}_1 \rho_1 {\mathbf \nabla}\cdot{\mathbf
                       J}_1 .
\end{equation}
In the limit of spherical symmetry, the vector
part  ${\mathbf J}_{t} \equiv J_t (r) {\mathbf e}_{r}$ is the only
non-vanishing part of the tensor density ${\mathbb J}_{\mu\nu}$.
Hence, in this limit, the tensor part of the functional (\ref{htta})
reduces to:
\begin{equation} \label{h_tens_spher}
\mathcal{H}^T     = \frac{1}{2}C^J_0
{\mathbf J}^2_0 +
                       \frac{1}{2}C^J_1 {\mathbf J}^2_1.
\end{equation}
By performing variation of the functional with respect
to $J_t (r)$ one obtains the one-body SO potential:
\begin{eqnarray}\label{sot} W_t^{SO} &
= & - \frac{1}{2r}\left( C^{\nabla J}_t \frac{d\rho_t}{dr} - C^J_t J_t(r)\right)
{\mathbf L} \cdot {\mathbf S},
\end{eqnarray}
which is composed of two terms. The first term is
coming form the SO term in the functional, Eq.~(\ref{httb}). It is proportional
to the radial derivative of the particle density and is relatively
slowly varying with $N$ and $Z$. The second component is due to the
tensor term (\ref{htta}). It is proportional to the SO density $J_t (r)$ which
is strongly shell-filling dependent. Indeed,
in the spherical symmetry limit, the SO vector
density can be written as~\cite{[Vau72]}:
\begin{eqnarray}
J(r) &=& \frac{1}{4\pi
r^3}\sum_{n,j,l}(2j+1)v^2_{njl} \nonumber \\ \label{sov_den}
     & & \times \left[ j(j+1)-l(l+1)-\frac{3}{4} \right] \psi^2_{njl}(r),
\end{eqnarray}
where $v^2_{njl}$ and $\psi^2_{njl}(r)$ are occupation probabilities and
radial wave functions, respectively, of states with given quantum numbers.
If both SO partners $j_\gtrless=l \pm 1/2$ are fully occupied,
i.e., when the system is spin-saturated (SS) the $J(r)$ density vanishes.
Examples of the SS systems include  $^{16}$O,
$^{40}$Ca, or $^{80}$Zr at spherical shape. Most of the nuclei are
spin-unsaturated (SUS). The SO vector density reaches its maximum when one
(or more) of the SO partners is fully occupied while the other one is
completely empty.

\vspace{0.5cm}

\section{Fitting procedure}
\label{sec3}

Fitting procedure used to constrain the coupling constants
$C^J_t$ and $C^{\nabla J}_t$ was described in detail in Ref.~\cite{[Zal08]}
and we only recall it here very briefly.
The idea is to reproduce experimental $f_{7/2} - f_{5/2}$ SO splittings in
three key nuclei: $^{40}$Ca, $^{48}$Ca, and $^{56}$Ni.
Since $^{40}$Ca is, as discussed above, a SS system, the
conventional SO term of Eq.~(\ref{httb}) is the only source of the SO
splitting. Hence, this nucleus is used to set the isoscalar strength of
the SO term $C^{\nabla J}_0$.
Having set $C^{\nabla J}_0$, we next constrain the $C^J_0$ strength by using
the $f_{7/2} - f_{5/2}$ SO splitting in the isoscalar ($N=Z$)
SUS nucleus $^{56}$Ni. Finally, we move to
$^{48}$Ca, where protons and neutrons constitute a SS and SUS system, respectively.
This nucleus is used to fit the isovector coupling constants or, more precisely,
to fit $C^{\nabla J}_1$, because the ratio $C^{\nabla J}_0/C^{\nabla J}_1$ is
kept equal to the value characteristic for
the given Skyrme parametrization. There is
one piece of data on the $f_{7/2} - f_{5/2}$ SO splittings,
preferably in $^{48}$Ni or $^{78}$Ni,
which is badly needed to fit the tensor and SO terms uniquely.

\begin{figure}[!]
\includegraphics[width=0.4\textwidth, clip]{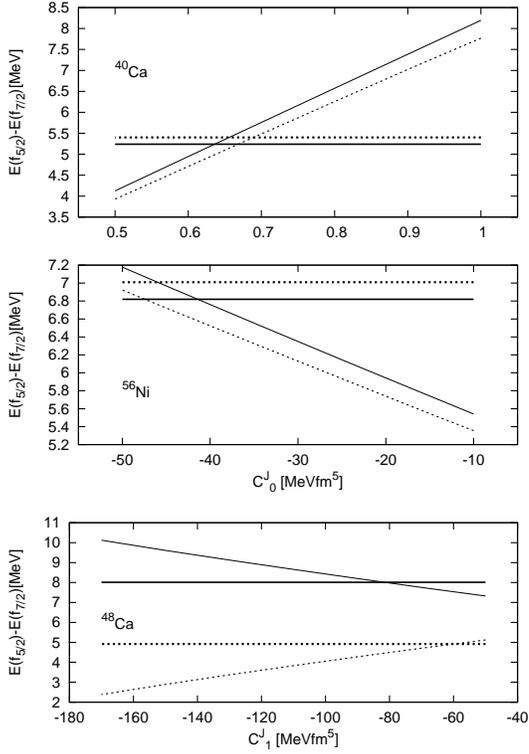}
\caption{Empirical (horizontal lines) and theoretical (inclined lines)
$f_{7/2} - f_{5/2}$ splittings in $^{40}$Ca, $^{48}$Ca, and $^{56}$Ni
as a function of $C_0^{\nabla J} / C_0^{\nabla J}(\text{SLy4})$ (upper panel),
$C_0^J$ (middle panel), and $C_1^J$ (lower panel). Solid an dashed
lines represent neutron an proton splittings, respectively.
The theoretical results are obtained by modifying the SO and tensor
strengths in the SLy4 functional.
Empirical data are taken from~\cite{[Oro96w]}. See text for further
details.}
\label{fitting} \end{figure}

The procedure outlined above is illustrated in Fig.~\ref{fitting}
for the case of the SLy4 functional~\cite{[Cha97fw]} but it is
qualitatively independent of the initial parameterization. As shown in the
figure, a good agreement with empirical data requires, for this
low-effective-mass force, circa 35\% reduction of $C_0^{\nabla J}$ as
compared to the original value, a large attractive isoscalar tensor coupling
constant $C_0^{\nabla J}$ of about
$-45$\,MeV\,fm$^5$, and $C_1^{\nabla J}$ of about $-70$\,MeV\,fm$^5$.
It appears that the resulting tensor coupling
constants $C_t^{\nabla J}$ (as well as the SO strengths $C_0^{J}$) are, to
large extent, independent on the initial parameterization. This is illustrated
in Fig.~\ref{fig1} where different functionals modified according to our
prescription, see Refs.~\cite{[Zal08],[Sat08],[Zal08as]}, are collected. They
are labeled by a subscript $T$ following the force acronym and marked
by black triangles.

\vspace{0.5cm}

\section{Contribution from the tensor terms to the binding energy}
\label{sec4}

\begin{figure}[t]
\includegraphics[width=0.48\textwidth, clip]{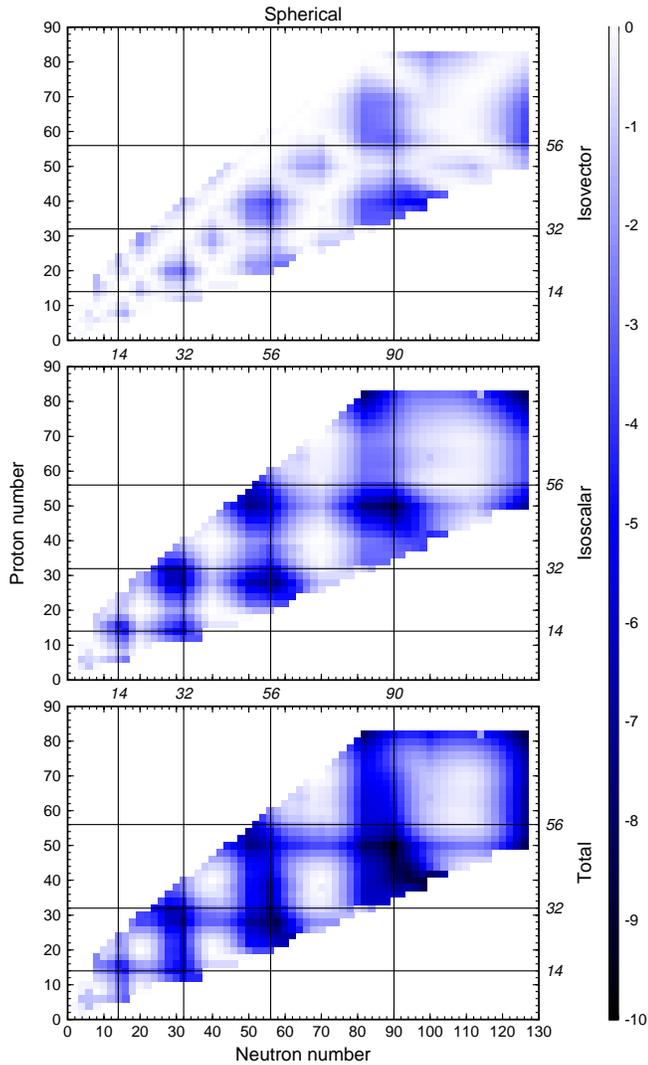}
\caption{The isovector (upper), the isoscalar (middle) and the total
(lower) tensor
contribution to the nuclear binding energy. The calculations were done using
the SLy4$_T$ interaction in the particle-hole channel and the volume-$\delta$
interaction in the particle-particle channel with spherical symmetry assumed.
Vertical and horizontal lines indicate the single-particle tensorial magic numbers
at spherical shape. See text for further details. } \label{hfb}
\end{figure}

Fig.~\ref{hfb} shows the contribution to the total nuclear binding
energy due to the tensor term, calculated by
using the spherical Hatree-Fock-Bogolubov (HFB)
code HFBRAD ~\cite{[Ben05]} with the SLy4$_T$ functional.
Contributions due to the
isovector and isoscalar parts are depicted separately in the upper and middle
panels, respectively. The total contribution, $B_T (N,Z)$, is shown in the lowest
panel of the figure.

From these results one can see that the isovector component is rather weak.
Hence, the topology of the total contribution to the energy is mostly determined
by the isoscalar term that shows a strong shell dependence. Following the
argumentation presented in Sec.~\ref{sec2}, the strongest
tensor effects are expected to appear for $N(Z)$=14, 32,
56, and 90. They correspond to nucleons filling the
$1d_{5/2}$, $1f_{7/2}\oplus 2p_{3/2}$, $1g_{9/2}\oplus 2d_{5/2}$ and
$1h_{11/2}\oplus 2f_{7/2}$ shells, respectively, which creates a maximum
SUS filling. Hence, these numbers can be regarded as
{\textit {tensorial magic numbers}}.

However, as it is seen from Fig.(\ref{hfb}), the $B_T (N,Z)$
does not follow the expected pattern exactly.
This is due to ({\it i}) pairing-induced configuration mixing and
({\it ii}) changes in the s.p.\ ordering of levels caused by
the combination of strong attractive tensor fields and strongly reduced
SO field. Two such situations are visible in Fig.~\ref{hfb}. For $N < 30$
the tensor contribution is, as expected, largest for $_{32}$Ge.
For $40 < N < 50$, however, the minimum on the
plot is shifted toward the $_{28}$Ni isotopes,
which suggests a change in the order of the $1p_{3/2}$ and $1f_{5/2}$ proton sub-shells
with increasing neutron excess. The figure also indicates
that on the proton side, $Z = 50$ rather than $Z = 56$ is the tensorial magic number.
Again, this suggests that the $1g_{7/2}$ proton sub-shell is
filled before $1d_{5/2}$. Consequently, the tensorial magic numbers may
slightly differ for neutrons ($N$ = 14, 32, and 56) and for protons
($Z$ = 14, 28, and 50).  This effect, however,
may strongly depend upon a rather delicate balance between the tensor and SO
strengths and needs to be studied in detail.

\section{Summary}\label{summary}

A new strategy of fitting the coupling constants of the nuclear
energy density functional was recently proposed by our group~\cite{[Zal08]}.
It is based  on a fit of the isoscalar spin-orbit
and both isoscalar and isovector tensor coupling constants directly to
the the $f_{5/2}-f_{7/2}$ SO splittings in $^{40}$Ca, $^{56}$Ni, and $^{48}$Ca.
Our results show that drastic changes in the isoscalar SO strength
and the tensor coupling constants are required as compared
to the commonly accepted values

This work briefly addresses the consequences of strong
attractive tensor and weak SO fields on binding energies.
In particular, a contribution to the binding energy
due to the tensor interaction is calculated. It shows a generic
pattern closely resembling that of the shell-correction. The {\it
tensorial magic numbers\/} are shifted up relatively to the standard magic numbers,
towards $N(Z)$=14, 32, 56, or 90,
which, in the extreme s.p.\ scenario at the spherical shape,
correspond to the maximum spin-orbit asymmetry in the
$1d_{5/2}$, $1f_{7/2}\oplus 2p_{3/2}$,  $1g_{9/2}\oplus 2d_{5/2}$,
and $1h_{11/2}\oplus 2f_{7/2}$
configurations, respectively.
It is shown that these numbers are
smeared out by the pairing effects and shifted in the case of protonic
tensorial magic
numbers by changes in the sub-shell ordering.
It is also shown that strong attractive tensor interaction may
give rise to an increased stability of nuclear binding at the drip lines, in
particular around $Z\approx 14, N\approx 32$,
$Z\approx 28, N\approx 56$, and  $Z\approx 50, N\approx 90$.

\vspace{0.3cm}

This work was supported in part by the Polish Ministry of
Science under Contract No.~N~N202~328234,
by the Academy of Finland and University of
Jyv\"askyl\"a within the FIDIPRO programme, and by the
Swedish Research Council.


\end{document}